%% file: main.tex
\documentclass[submission,copyright,creativecommons]{eptcs}
\usepackage{breakurl}             
\usepackage{amstext}
\usepackage{amssymb}
\usepackage{amsmath}
\usepackage{xcolor}
\usepackage{stmaryrd}
\usepackage{proof}
\usepackage{alltt}
\usepackage{graphicx}
\usepackage{tikz}

\title{An Introduction to Liquid Haskell}
\author{Ricardo~Pe\~na
\institute{Complutense University of Madrid\thanks{Work partially funded by the Spanish Ministry of Economy and Competitiveness, under the grant TIN2013-44742-C4-3-R},~
Spain}
\email{ricardo@sip.ucm.es}
}
\def\titlerunning{An Introduction to Liquid Haskell}
\def\authorrunning{R.~Pe\~na}

\newcommand{\imp}{\rightarrow}
\newcommand{\Sem}[1]{\mbox{$\llbracket #1\rrbracket$\/}}
\newcommand{\Imp}{\Rightarrow}
\newcommand{\len}[1]{\mathit{len}~#1}

\newcommand{\bool}{\ensuremath{\mathit{bool}}}
\newcommand{\integer}{\ensuremath{\mathit{int}}}

\begin{document}
\maketitle

\begin{abstract}
This paper is a tutorial introducing the underlying technology and the use of the tool Liquid Haskell, a type-checker for the functional language Haskell that can help programmers to verify non-trivial properties of their programs with a low effort.

The first sections introduce the technology of Liquid Types by explaining its principles and summarizing how its type inference algorithm manages to prove properties. The remaining sections present a selection of Haskell examples and show the kind of properties that can be proved with the system. 
\end{abstract}

\input{introduction}

\input{liquidtypes}

\input{liquidhaskell}

\input{totality}

\input{sorted}

\input{avl}

\input{conclu}


\input{main.bbl}
\end{document}

%% file: introduction.tex
\section{Introduction}
\label{sc:introduction}

This tutorial aims at exposing the reader to a first contact with the Liquid Types technology \cite{DBLP:conf/pldi/RondonKJ08}, and in particular with its application to the language Haskell, a type-checker known as {\it Liquid Haskell} \cite{DBLP:conf/haskell/VazouSJ14,DBLP:conf/icfp/VazouSJVJ14}. In the view of this author, Liquid Types should be regarded as a computer-assisted verification system that may increase the reliability of programs by paying a fraction of the effort needed by doing formal verification by hand.

Program verification is as old as programming. The first formal reasoning about programs was proposed by Alan Turing in 1949 \cite{Turing49}. The first set of axioms for a high-level programming language, was given by Tony Hoare in 1969 \cite{Hoa.69}. After that, the decade of 1970 saw the publication of plenty of papers and books about formal verification, formal derivation, and of all kind of proposals for applying formal methods and mathematical logic to reasoning about program correctness.

In spite of such a big effort, and of the fact that many universities include formal program verification in their curricula, forty years after we should admit that formal verification is far from being part of day-to-day programming. There are a number of reasons for this situation:

\begin{itemize}
\item It takes some effort to formalise the specification of methods by writing a precondition and a postcondition for each of them.

\item It takes much more effort to guess the loop invariants, and other critical intermediate assertions of programs.

\item Even having written all the critical assertions, writing and proving by hand all the verification conditions, need writing a text between 5 to 10 times the volume of the code being verified.

\end{itemize}
The general impression is then than formal verification gives us obvious benefits, but the effort investment needed to get them is too high. As a consequence, formal methods are barely used, and only in a few safety critical systems such an investment seems to be justified.

Between the two extremes of not verifying anything, or verifying every sentence of a program, some intermediate scenarios have been tried. Many programming systems  (e.g.\ \cite{Chalin05}) offer the possibility of including assertions in programs, and optionally executing them. This facility is equivalent to doing testing while the system is in operation, and may capture some bugs at a low cost. Sometimes, they even try to prove the assertions statically (e.g.\ \cite{Chalin05,Burdy05}). The kind of properties they prove are usually simple ones, such as detecting null pointer dereferencing, or array indices out of bounds, but again some bugs can be captured with a low effort investment.

Liquid Types have managed to find a way of getting many of the benefits of doing formal verification without paying much of the cost:

\begin{enumerate}
\item They usually require the user to give the precondition and the postcondition of functions, although in simple cases they can even infer them automatically.

\item They do not require to give the loop invariants or other intermediate assertions. The system can usually infer them.

\item The verification conditions which must hold for the program to be correct, are automatically extracted and proved by the system.

\end{enumerate}
Their main limitation is the kind of properties which can be proved in this way. Their formulas must belong to a decidable logic, as they are the logics supported by the SMT provers \cite{DBLP:conf/tacas/MouraB08,DBLP:conf/cav/BarrettCDHJKRT11}, which are the underlying proving machinery of Liquid Types.
These tools have evolved very quickly in the last ten years and currently they can deal with formulas including all the logical connectives, some of them even with existential and universal quantification, and the formulas also support integer and real linear arithmetic, algebraic types, arrays, and uninterpreted functions.

Liquid Types do not generate quantified formulas. Even though, it is surprising the broad spectrum of properties they can express and prove, as we will try to show in this tutorial. They include the automatic verification of many well-known sorting algorithms, and the preservation of the AVL-tree invariant by their associated operations.

The Liquid Types were originally developed in a functional language framework, and later on they were applied to some imperative languages such as C \cite{DBLP:conf/cav/RondonBKJ12}. Recently, they have been incorporated to Haskell \cite{DBLP:conf/haskell/VazouSJ14,DBLP:conf/icfp/VazouSJVJ14} in the form of a static type-checker which is independent of the compilers. The Hindley-Milner Haskell type system, and its extension to type classes, combine very well with the Liquid Types approach, which supports polymorphism, algebraic types, and lambda abstractions. Recently, even monad support has been incorporated to Liquid Haskell.

%% file: liquidtypes.tex
\section{Liquid Types }
\label{sc:liquidtypes}

Liquid types, an abbreviation of {\it Logically Qualified Data Types}, were first introduced in
\cite{DBLP:conf/pldi/RondonKJ08}. They were presented as a ``combination of Hindley-Milner type inference with Predicate Abstraction to automatically infer dependent types precise enough to prove a variety of safety properties''. Behind this definition there are different techniques:

\begin{itemize}
\item The Hindley-Milner type inference algorithm, usually associated to modern functional languages. This is not strictly essential to the approach. Liquid types could be equally applied to programming languages having a variety of type systems.

\item Predicate abstraction \cite{DBLP:conf/cav/GrafS97,DBLP:conf/pldi/SrivastavaG09}. This is a technique based on abstract interpretation which searches for the strongest predicate satisfying a set of constraints in a finite complete lattice of predicates related by an entailment relation. This is an essential part of the liquid type approach.

\item Dependent types \cite{DBLP:conf/icfp/Augustsson98}. These are types that depend on the values computed by the program. It can be used to express program properties holding in different parts of a program text. In their full generality, the corresponding type inference problem is undecidable and so heavy manual intervention by the programmer is needed. Liquid types are a restricted version of dependent types in which type inference is decidable.

\end{itemize}

A liquid type has the form $\{\nu:\tau\mid e\}$, where $\tau$ is a Hindley-Milner type and $e$ is a boolean expression which may contain the $\nu$ variable and free variables occurring in the program. This type represents all the values $u$ of type $\tau$ such that the expression $e[u/\nu]$ evaluates to \emph{true}. For instance, the type $\{\nu:\integer \mid \nu > x\}$ represents the type of all the integers greater than the value of the free variable $x$. This is called a {\it refinement type} of the type $\integer$. The $\nu$ is called the {\it value variable}, and it is assumed to range over the values of the refinement type.

In a function definition, the type of the result is allowed to depend on the value of the arguments. Moreover, the type of an argument can depend on the value of a preceding argument. For instance, the following function receives a polymorphic array, an index within the expected range, and gets the array element at that position:

\begin{eqnarray*}
\texttt{get} & :: &  \forall \alpha.(a:\mathit{array}\ \alpha) 
    \rightarrow i : \{ \nu : \integer \mid 0 \leq \nu < \len{a}\}
    \rightarrow \{ \nu : \alpha \mid \nu = a[i] \} 
\end{eqnarray*}

The programming language is given a set of typing rules expressing the relationships that must hold between the liquid types in order that the program is well-typed. The most important one is that of subtyping: intuitively, a liquid type $\tau_1$ is a subtype of a liquid type $\tau_2$, expressed $\tau_1<:\tau_2$,  if the set of values of $\tau_1$ is a subset of the set of values of $\tau_2$. In logical terms, if $\tau_1=\{\nu:\tau\mid e_1\}$ and $\tau_2=\{\nu:\tau\mid e_2\}$, this is equivalent to show that the formula $e_1\Imp e_2$ is universally valid. But the typing rules do it better: they collect in the typing environment $\Gamma$, not only the types of all the free variables in scope, as usual, but also the boolean conditions that hold at the text location where the subtype relation is proved. These conditions are collected from the boolean discriminants of the {\bf if} expressions. Then, the formula that must be shown valid is $\Sem{\Gamma}\wedge e_1\Imp e_2$, where $\Sem{\Gamma}$ contains these conditions, and also the liquid types of the variables in scope converted into boolean expressions, i.e.\ each binding of the form $x:\{\nu:\tau \mid e\}$ is translated into the boolean formula $e[\nu/x]$.

\begin{figure}
\[
\begin{array}{c}
\infer[WEAK]{\Gamma\vdash e:t_2}{\Gamma\vdash e:t_1\qquad \Gamma\vdash t_1<: t_2}
\qquad\qquad
\infer[VAR]{\Gamma\vdash x:\{\nu: B\mid x=\nu\}}{\Gamma(x)=\{\nu :B\mid e\}}
\\[1em]
\infer[SUBTYPE]{\Gamma\vdash \{\nu: B\mid e_1\}<:\{\nu: B\mid e_2\}}
{\mathit{valid}(\Sem{\Gamma}\wedge e_1\Imp e_2)}
\\[1em]
\infer[IF]{\Gamma\vdash\mathbf{if}~ e_1~ \mathbf{then}~ e_2 ~\mathbf{else}~ e_3: t}
{\Gamma\vdash e_1:\bool\qquad\Gamma;e_1\vdash e_2:t\qquad\Gamma;\neg e_1\vdash e_3:t}
\end{array}
\]
\caption{Some liquid typing rules}
\label{fg:rules}
\end{figure}

In order to clarify the kind of formulas that the system must prove valid, we show in Fig.~\ref{fg:rules} some typing rules taken from \cite{DBLP:conf/pldi/RondonKJ08}. There, $B$ represents a non-functional basic type, and $t, t_1,\ldots$ represent arbitrary liquid types. By using those rules, we are proving correct a {\it max} function computing the maximum of two values, having the following specification
\[
\mathit{max}: x:\integer\imp y:\integer\imp \{\nu:\integer\mid \nu\ge x\wedge \nu\ge y\}
\]
and the following code:
\[
\mathit{max}~x~y=\mathbf{if}~x\ge y~ \mathbf{then}~x~\mathbf{else}~y
\]
The type derivation collects the following sequence of proof obligations:
\[
\begin{array}{ll}
\mbox{by the IF rule} & [x:\integer,~y:\integer]\vdash x\ge y:\bool\\
\mbox{by the IF rule} & [x:\integer,~y:\integer,~x\ge y]\vdash x: \{\nu:\integer\mid \nu\ge x\wedge \nu\ge y\}\\
\mbox{by the IF rule} & [x:\integer,~y:\integer,~\neg(x\ge y)]\vdash y: \{\nu:\integer\mid \nu\ge x\wedge \nu\ge y\}\\
\mbox{by the WEAK, VAR and SUBTYPE rules}  & \mathit{valid}(\nu =x\wedge x\ge y\Imp \nu\ge x\wedge \nu\ge y)\\
\mbox{by the  WEAK, VAR and SUBTYPE rules} & \mathit{valid}(\nu =y\wedge \neg(x\ge y)\Imp \nu\ge x\wedge \nu\ge y)\\
\end{array}
\]
Obviously, the typing $[x:\integer,~y:\integer]\vdash x\ge y:\bool$ is correct, and the last two formulas are universally valid, so the function {\it max} type-checks.

Should the user provide the liquid types of {\it all} the program variables, then pure type checking would consist of finding a type derivation for the program by applying the typing rules of the language, and then discharging all the proof obligations coming from the subtyping relations. In order to prove all the formulas automatically, a first requirement of the Liquid Type System (LTS) is that they must belong to a decidable logic. Assuming this, then the system uses an SMT solver to discharge the validity of the formulas.

But, annotating by hand the liquid types of all the variables would be a heavy burden for the programmer. Fortunately, the LTS requires a minimum hand annotation. In most  cases, only the type signature of the function being proved is required, i.e.\ the types of the arguments and that of the function result. In this signature, the user must express the dependence between arguments, and also how the result depends on the arguments. This amounts to giving the function precondition and postcondition, i.e.\	 its specification, said in classical program verification terms.

With this information, the LTS tries to infer the liquid types of all the intermediate program variables and program subexpressions. This would be a hopeless search if no restrictions were posed to the shape of the predicates that may occur in the types. To this aim, the following restrictions are posed:

\begin{itemize}
\item The predicates $e$ occurring in liquid types of the form $\{\nu :t\mid e\}$ are restricted to be {\it conjunctions} of atomic qualifiers $q$ belonging to a set $\mathbb{Q^*}$.

\item The set $\mathbb{Q^*}$ is different at each text location. All the sets $\mathbb{Q^*}$ are obtained from an only set $\mathbb{Q}$ given by the programmer, by substituting variables in scope at the corresponding text location for all the occurrences of the wildcard symbol $\star$ in $\mathbb{Q}$.

\item After that, all the ill-typed qualifiers are removed. Only well-typed ones remain in each $\mathbb{Q^*}$.

\end{itemize}
For instance, assuming that $\nu$ ranges over the type \integer, if $\mathbb{Q}=\{\nu \ge 0, \star \le \nu, \nu < \len{\star}\}$, and the variables in scope are two integer numbers $x,~y$, and an array  $a$, then 
$\mathbb{Q^\star}=\{\nu \ge 0, x\le \nu,y\le \nu,\nu < \len{a}\}$. Qualifiers such as $a\le \nu, \nu<\len{x}$ will be generated and then removed for being ill-typed.

These restrictions ensure that the number of candidate predicates at each program location is finite, so an exhaustive search would do the job by trying, for the (finite) set of program locations, all possible combinations of predicates. If a combination makes all the proof obligations valid, then the program type-checks. This brute-force approach is unpractical for even very small programs. The LTS does it better by organizing the search in a complete lattice of predicates, and then going from the strongest possible predicate to weaker ones upwards in this lattice. At each step the {\it weakening} of a single predicate, i.e.\ of a single liquid type, is done in order to make a proof obligation valid. If a solution is found, it is guaranteed that it is the strongest posssible one. This amounts to saying that the smallest types have been found for every program variable and program subexpression.

Of course, even if the program is correct, a solution may not be found for a number of reasons:

\begin{itemize}
\item The set of qualifiers given  in  $\mathbb{Q}$ by the programmer is not enough. A solution exists if additional qualifiers were included in  $\mathbb{Q}$.

\item Even if $\mathbb{Q}$ is big enough, the solution may include some disjuntions of the given qualifiers, and this is not allowed by the approach.

\end{itemize}

But, if a typing exists with the given $\mathbb{Q}$, and the restriction of liquid types being conjunctions of qualifiers, then the system is guaranteed to find it.

%% file: liquidhaskell.tex
\section{Liquid Haskell}
\label{sc:liquidhaskell}

Liquid Haskell (LH) was first introduced in \cite{DBLP:conf/haskell/VazouSJ14,DBLP:conf/icfp/VazouSJVJ14}. It represents the application of the Liquid Type theory to a full-fledged functional language like Haskell. It consists of a static type-checker for a big part of the Haskell language. The first phase of LH uses the Haskell compiler GHC \cite{DBLP:conf/fp/HallHPJW92} in order to solve the external references, to type-check the program in the Hindley-Milner sense, and to transform it to its internal Core representation. This transformation simplifies the work of LH,  since it then only deals  with a few syntactic constructions.

The Liquid type annotations are provided by the programmer in the input file as Haskell comments of the form \texttt{\small \{-@ annotation @-\}}. These, of course, are ignored by GHC and are instead processed by LH. As a result, a set of type constraints are generated in the second phase, which are solved in a third phase with the help of a SMT solver, such as Z3 \cite{DBLP:conf/tacas/MouraB08} or CVC4 \cite{DBLP:conf/cav/BarrettCDHJKRT11}. The input file also contains the set of qualifier fragments from which the inferred liquid types are to be built. Due to a judicious choice of defaults, by which the qualifier fragments are directly extracted from the type annotations, this set is most of the times empty. 

The output of LH is a simple word \texttt{SAFE}, in the case that every function in the input file type-checks. Otherwise, type errors are reported at different text locations, indicating the inferred types and the constraints which have been violated. The error reports are usually informative enough to detect and repair the problem. They constitute a big help for debugging the program.

In order to install LH and to get a complete tutorial with exercises, visit the following pages;

\begin{eqnarray}
\texttt{https://github.com/ucsd-progsys/liquidhaskell}\\
\texttt{http://ucsd-progsys.github.io/liquidhaskell-tutorial/} \label{eq:tutorial}
\end{eqnarray}

You will need a Haskell installation, and also to install Z3, CVC4, or other SMTLIB compatible solver.

Liquid Haskell has been applied to over 10.000 lines of Haskell code belonging to different popular libraries, as reported in \cite{DBLP:conf/haskell/VazouSJ14}. The properties specified and proved range from totality and termination of functions, to safe access of indexed structures, and preservation of data type invariants. In sections \ref{sc:sorted}  and \ref{sc:avl} we show a selection of case studies taken from \cite{DBLP:conf/haskell/VazouSJ14} and from the above cited tutorial.

We finish this section by enumerating some of the annotations that a programmer may include in a LH input file:
\begin{description}
\item[\texttt{type}] This allows to define an alias for a liquid type. The definition may include as arguments type variables (in lower case) and value variables (in upper case).

\item[\texttt{data}] Similar to the {\bf data} declaration of Haskell to introduce algebraic types, but here the programmer may indicate that the type of a constructor argument depends on the value of prior argument.

\item[\texttt{measure}] This annotation specifies the name of a Haskell function as a {\it measure}. A measure is a possibly recursive function which can be used in type definitions. Examples of measures are the length of a list, or the height of a tree. We will give examples using measures in sections~\ref{sc:sorted} and \ref{sc:avl}.

\item[\it Function signature] This allows to give a liquid type to a function. Its definition will be normal Haskell code.

\end{description}

%% file: totality.tex
\section{Totality}
\label{sc:totality}

Very frequently in Haskell, we define partial functions such as the one getting the head element of a list:

\begin{alltt}\small
  head :: [a] -> a
  head (x:_) = x
\end{alltt}
The translation of this definition made by GHC is:
\begin{alltt}\small
  head y = {\bf case} y {\bf of}
              x:_ -> x
              []  -> patError "head"
\end{alltt}
Very frequently also, we get runtime errors when a part of the program is calling a partial function outside of its definition domain:

\begin{alltt}\small
  *** Exception: Prelude.head: empty list
\end{alltt}

LH can help us to verify at compile time that this kind of errors will not happen at runtime. The first thing to do is defining a so-called {\it boolean measure}:

\begin{alltt}\small
  \{-@ measure notEmpty @-\}
  notEmpty :: [a] -> Bool
  notEmpty []    = False
  notEmpty (_:_) = True
\end{alltt}

Measures are Haskell total functions having a very restricted syntax, that LH converts into uninterpreted functions of the underlying SMT theory satisfying a set of axioms. It gives the following types to the list constructors:
\begin{alltt}\small
  []  :: \{v: [a] | notEmpty v = False\}
  (:) :: a -> [a] -> \{v: [a] | notEmpty v = True\}
\end{alltt}
After that, we strengthen the signature of \texttt{head} by giving it the following type:
\begin{alltt}
  \{-@ type NEList a = \{v:[a] | notEmpty v\} 
      head :: NEList a -> a  @-\}
\end{alltt}
LH succeeds in type-checking the above definition for \texttt{head}. To verify its (Core) definition, LH checks the body expression with a $\Gamma$ typing environment having the restriction \texttt{notEmpty y}. The first {\bf case} branch succeeds, since \texttt{y} is matched with \texttt{x:\_} and then \texttt{notEmpty y} holds. In the second {\bf case} branch \texttt{y} is matched with \texttt{[]}, and then we get the contradiction:
\begin{alltt}\small
  y :: notEmpty y && not (notEmpty y)
\end{alltt}
A type refinement {\tt False} is an unhabited type. So, LH concludes that the call to \texttt{patError} is dead code, and this confirms the totality of head.

With this type, now the burden is on the side of the users of \texttt{head}.
For every call to it, LH must ensure that the list passed as an argument is in fact non-empty. If it succeeds in this checking, then the above pattern error will never happen at runtime.

%% file: sorted.tex
\section{Case Study: sorted lists}
\label{sc:sorted}

In \cite{DBLP:conf/pldi/KawaguchiRJ09}, the original idea of liquid types is extended to recursive algebraic datatypes, giving rise to the so-called {\it recursive refinements}. There, the type of an argument of a data constructor may depend on the value of a prior argument. Together with a recursive definition, this feature is powerful enough to allow defining interesting invariants of data structures such as sortedness of lists:

\begin{alltt}\small
  \{-@ data IncList a = Emp
                     | (:<) \{ hd::a, tl::IncList \{v:a | hd <= v\}\} @-\}
\end{alltt}
Here, an increasing sorted list is defined by restricting the list tail elements to be not smaller than the head. This property is recursively propagated to all the sublists.
LH interprets this definition by assigning the following types to the data constructors:

\begin{alltt}\small
  Emp  :: IncList a
  (:<) :: hd:a -> tl:IncList \{v:a | hd <= v\} -> IncList a
\end{alltt}
Given this invariant, and the appropriate signature for a function \texttt{insert} inserting an element in a sorted list, LH is able to type-check the following definition:

\begin{alltt}\small
  insert :: (Ord a) => a -> IncList a -> IncList a
  insert y Emp                   = y :< Emp
  insert y (x :< xs) | y <= x    = y :< x :< xs
                     | otherwise = x :< insert y xs
\end{alltt}
Notice in the last line that LH needs to infer the type \texttt{IncList (\{v:a | a <= x\})} for the subexpression \texttt{insert y xs} in order this equation to type-check, which is far from being trivial. Using the signature just proved for \texttt{insert}, it is less surprising that LH also type-checks the following code for the insertion sort algorithm:

\begin{alltt}\small
  insertSort :: (Ord a) => [a] -> IncList a
  insertSort []     = Emp
  insertSort (x:xs) = insert x (insertSort xs)
\end{alltt}

Similarly, we can give the following signature and code of a function merging two sorted lists into a single sorted one:
\begin{alltt}\small
  merge :: (Ord a) => IncList a -> IncList a -> IncList a
  merge xs        Emp       = xs
  merge Emp       ys        = ys
  merge (x :< xs) (y :< ys)
    | x <= y                = x :< merge xs (y :< ys)
    | otherwise             = y :< merge (x :< xs) ys
\end{alltt}
LH is able to type-check this definition and, again, the types inferred for the subexpressions \texttt{merge xs (y :< ys)} and \texttt{merge (x :< xs) ys} are far from being trivial. In the first case, it is \texttt{IncList (\{v:a | a <= x\})}, and in the second one it is \texttt{IncList (\{v:a | a <= y\})}. Assuming implemented a function \texttt{split::[a] -> ([a],[a])} splitting a list into two, LH successfully type-checks the following signature and code for the {\it mergesort} algorithm:
\begin{alltt}\small
  mergeSort :: (Ord a) => [a] -> IncList a
  mergeSort []     = Emp
  mergeSort [x]    = x :< Emp
  mergeSort xs     = merge (mergeSort ys) (mergeSort zs)
    where (ys, zs) = split xs
\end{alltt}

If the types provided by the programmer  in the signatures are not strong enough, then LH will complain, and the error reports may help him/her to repair the problem. Let us assume the following definition for the {\it quicksort}  algorithm:

\begin{alltt}\small
  quickSort :: (Ord a) => [a] -> IncList a
  quickSort []     = Emp
  quickSort (x:xs) = join x lessers greaters
    where lessers  = quickSort [y | y <- xs, y < x ]
          greaters = quickSort [z | z <- xs, z >= x]

  join :: a -> IncList a -> IncList a -> IncList a
  join z Emp       ys = z :< ys
  join z (x :< xs) ys = x :< join z xs ys
\end{alltt}
LH will complain about the type given to \texttt{join}. It cannot deduce that the result list is sorted  from just the fact that the two input lists are sorted. It would need stronger types for the two input lists. After some trial and error, the programmer would eventually arrive at the following correct type:
\begin{alltt}\small
  join :: x:a -> IncList (\{v:a | v <= x\}) -> IncList (\{v:a | x <= v\}) -> IncList a
\end{alltt}
Notice, again, that in order to type-check the second equation of \texttt{join}, LH infers for the subexpression \texttt{join z xs ys} the type \texttt{IncList (\{v:a | x <= v\})}, which is a refinement of the result type \texttt{IncList a}.

%% file: avl.tex
\section{Case Study: AVL trees}
\label{sc:avl}

Another important data structure whose invariant can be elegantly expressed with liquid types are {\it binary search trees}:

\begin{alltt}\small
  \{-@ data BST a = Leaf
                 | Node \{ root  :: a
                        , left  :: BST \{v:a | v < root \}
                        , right :: BST \{v:a | root < v \} \} @-\}
\end{alltt}
In this case, the recursive property is that, in every nonempty subtree, all the elements of the left child are smaller than the value at the root, and all those of the right child are greater than the root.

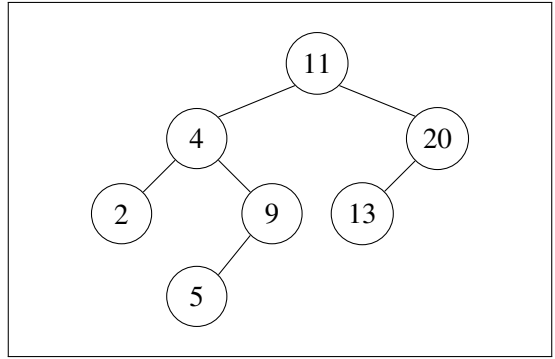
\begin{figure}
\begin{center}
\fbox{\parbox{7cm}{\begin{center}

\begin{tikzpicture}
[
roundnode/.style={circle, draw, 
minimum size=8mm}
]
\draw  (0,0)   node[roundnode]  (root)  {11};
\draw (root)+(-1.6,-1) node[roundnode]  (left-level1)  {4};
\draw (root)+(1.6,-1) node[roundnode]    (right-level1) {20};
\draw (left-level1)+(-1,-1) node[roundnode]   (left-level2) {2};
\draw (left-level1)+(1,-1) node[roundnode]  (center-level2) {9};
\draw (right-level1)+(-1,-1) node[roundnode] (right-level2) {13};
\draw (center-level2)+(-1,-1.1) node[roundnode]  (level3) {5};

\draw
(root.south west) -- (left-level1.north east);
\draw
(root.south east) -- (right-level1.north west);
\draw
(left-level1.south west) -- (left-level2.north east);
\draw
(left-level1.south east) -- (center-level2.north west);
\draw
(right-level1.south west) -- (right-level2.north east);
\draw
(center-level2.south west) -- (level3.north east);

\end{tikzpicture}
\end{center}
}}
\caption{An example of AVL tree}
\label{arboles.AVL.fg:AVL}
\end{center}
\end{figure}

But, in order to ensure a time cost in $O(\log n)$ for all the tree operations, AVL trees \cite[Chap. 10]{HSM.97}, have an invariant stronger than that of BST. In addition to that invariant, they require to be reasonably balanced, and by this it is meant that the difference of heights between the left and right children of every subtree is at most one. In Fig.~\ref{arboles.AVL.fg:AVL} we show an example of such a tree. We will define below a type \texttt{AVL} very similar to \texttt{BST} but keeping a new field in the nodes holding the height of the subtree having that node as its root. This will ease the checking of the balance property.

For expressing the latter, first we need to define a function \texttt{height} giving us the height of a tree and to inform LH that this is a {\it measure} for the type \texttt{AVL}:

\begin{alltt}\small
  \{-@ measure height @-\}
  \{-@ height :: AVL a -> Nat @-\}
  height Leaf = 0
  height (Node _ l r _) = 1 + max (height l) (height r)
\end{alltt}
where the type \texttt{Nat} has been previously defined as:
\begin{alltt}\small
  \{-@ type Nat = \{v:Int | 0 <= v\} @-\}
\end{alltt}

Measures may return any type (not only boolean values) but, as we have said, they have severe restrictions:
\begin{enumerate}
\item They must be total, and have exactly an equation per data constructor.
\item They may be recursive but the recursive function must be applied only to pattern variables. This ensures termination.
\item The terms in the righthand expressions must belong to the underlying SMT theory.
\end{enumerate}

In order to simplify the definition of the AVL type, we introduce the following declarations, taken from (\ref{eq:tutorial})\footnote{
This refers to the tutorial whose URL was given in Sec.~\ref{sc:liquidhaskell}.
}:

\begin{alltt}\small
  \{-@ type AVLL a X = AVL \{v:a | v < X\} @-\}
  \{-@ type AVLR a X = AVL \{v:a | X < v\} @-\}
  isReal h l r = h ==  nodeHeight l r
  nodeHeight l r = 1 + max (height l) (height r)
  isBal l r n = 0 - n <= d && d <= n              -- difference in height is at most n
        where d = height l - height r
\end{alltt}
The first type describes the AVLs that could be correctly installed as left children of a root with value $X$.  The second one is symmetrical for right children. The third definition is a predicate that the fourth component \texttt{h} of a node having \texttt{l} and \texttt{r} as children should meet: to exactly hold the height of the subtree having as its root such a node. The last one is a predicate expressing a balancing property between two subtrees \texttt{l} and \texttt{r}. The AVLs satisfy \texttt{isBal l r 1}.

Now, an AVL is a binary search tree that additionally is balanced:

\begin{alltt}\small
  \{-@ data AVL a = Leaf
                 | Node \{ key :: a
                        , l   :: AVLL a key
                        , r   :: \{v:AVLR a key | isBal l v 1\}
                        , ah  :: \{v:Nat        | isReal v l r\}
                        \}                                        @-\}
\end{alltt}

Let us do a first attempt of defining an \texttt{insert} function inserting an element into an AVL:

\begin{alltt}\small
  \{-@ insert :: (Ord a) => a -> AVL a -> AVL a @-\}
  insert y t@(Node x l r _) | y < x = node x (insert y l) r
                            | x < y = node x l (insert y r)
                            | otherwise = t
  insert y Leaf = Node y Leaf Leaf 0
\end{alltt}
where \texttt{node} is a {\it smart constructor}, taking care of not to violate the AVL invariant:

\begin{alltt}\small
  \{-@ node :: x:a -> l:AVLL a x -> r:\{v:AVLR a x | isBal l v 1\} 
                  -> AVLN a (nodeHeight l r)                      @-\}
  node x l r = Node x l r h
       where h       = 1 + max hl hr
             hl      = getHeight l
             hr      = getHeight r
\end{alltt}
Function \texttt{getHeight} just gets the height field of the root node, or returns 0 if it is an empty tree, and the auxiliary type \texttt{AVLN a H} defines the trees \texttt{AVL a} of height \texttt{H}:

\begin{alltt}\small
  \{-@ type AVLN a N = \{v: AVL a | height v = N\} @-\}
  \{-@ measure getHeight @-\}
  \{-@ getHeight :: t:AVL a -> \{v:Nat | v == height t\} @-\}

  getHeight Leaf           = 0
  getHeight (Node _ _ _ h) = h
\end{alltt}

The above definition for \texttt{insert} is wrong, and LH will complain that its result need not be an AVL. The obvious reason is that no effort has been done to preserve the balance property. As a consequence, unbalanced trees may be obtained. For instance, the term:

\begin{alltt}\small
  insert 3 (insert 2 (insert 1 Leaf)) 
\end{alltt}
will fail at the subexpression \texttt{\small node 1 Leaf (Node 2 Leaf (Node 3 Leaf Leaf 1) 2)} in which the smart constructor \texttt{node} refuses to join two trees with a height difference of 2.

By following the AVL version of \cite[Chap. 7]{Pen.05}, first we replace in the above definition for \texttt{insert} the smart constructor \texttt{node} by a smarter version \texttt{equil} that checks the height difference of the two children. Should this difference be at most one, then the constructor \texttt{node} would be called, as the AVL invariant would not be violated. The other possibility is the height difference to be exactly two. In that case, \texttt{equil} will decide whether the bigger tree is the left or the right one. In the first case, a function \texttt{leftUnbalance} will repair the unbalance by doing a left {\it rotation}. In the second case, a symmetric function \texttt{rightUnbalance} will be called. Let us see for the moment, the specification and the implementation of \texttt{equil}:

\begin{alltt}\small
  \{-@ equil :: x:a -> l:AVLL a x -> r:\{v:AVLR a x | isBal l v 2\} 
                   -> AVLE a \{height l\} \{height r\} @-\}

  equil x l r | isBal l r 1  = node x l r
              | hl == hr + 2 = leftUnbalance  x l r
              | hr == hl + 2 = rightUnbalance x l r

       where hl = getHeight l
             hr = getHeight r             
\end{alltt}
where the type \texttt{AVLE} expresses the expected height of the resulting AVL as a function of the input heights:
\begin{alltt}\small
  \{-@ type AVLE a H1 H2 = \{v: AVL a |                        
                        ((H2 <= H1 && H1 <= H2 +1)  => height v =  H1 + 1)
                     && ((H1 <= H2 && H2 <= H1 +1 ) => height v =  H2 + 1)
                     && (H1 = H2 + 2 => (H1 <= height v && height v <= H1 + 1))
                     && (H2 = H1 + 2 => (H2 <= height v && height v <= H2 + 1)) \} @-\}
\end{alltt}

The first two lines of the height property express what hapens when no rotation is needed, and then constructor \texttt{node} is invoked. The third line expresses what happens when a left rotation is needed, and the last one, is the symmetrical property for a right rotation.

This definition is type-checked by LH, provided the following signatures for \texttt{leftUnbalance} and \texttt{rightUnbalance} are given:

\begin{alltt}\small
  \{-@ leftUnbalance  :: x:a -> l:AVLL a x -> r:\{v:AVLR a x | height l == height r + 2\} 
                            -> AVLE a \{height l\} \{height r\} @-\}
  \{-@ rightUnbalance :: x:a -> l:AVLL a x -> r:\{v:AVLR a x | height r == height l + 2\} 
                            -> AVLE a \{height l\} \{height r\} @-\}
\end{alltt}

The code of \texttt{leftUnbalance} implements the so-called LL and LR rotations. In the first one, the unbalance is produced by the left child of the left child. Let us call it $\mathit{ll}$, and $\mathit{lr}$ to its sibling. If $h$ is the height of the right subtree $r$, then $h_\mathit{ll}=h+1$, and $h_\mathit{lr}=h$, or $h_\mathit{lr}=h+1$. By rearranging the subtrees in the order shown in the code below, the final tree will have a height $h+2$ in the first case, or $h+3$ in the second one, and it will satisfy the AVL invariant. 

In an LR rotation, the unbalance is produced by the right child of the left child, call it $\mathit{lr}$, which has a height $h+1$ (and then its parent has a height $h+2$), while the right subtree has a height $h$. This ensures that $\mathit{lr}$ is not empty, so it could be decomposed into its constituent pieces. By rearranging these pieces in the order shown in the code below, it is easy to check that the final tree is balanced, and it has a height $h+2$. The following code will be then type-checked by LH:

\begin{alltt}\small
  leftUnbalance x (Node y ll lr _) r
            | hll >= hlr = node y ll (node x lr r)
            | otherwise  = node z (node y ll lrl) (node x lrr r) 
 
     where hll              = getHeight ll
           hlr              = getHeight lr
           Node z lrl lrr _ = lr 
\end{alltt}
The code for \texttt{rightUnbalance} is symmetrical to that of \texttt{leftUnbalance}, and it is not shown.

By using the above smart constructor \texttt{equil}, the implementation of a function \texttt{delete}, removing an element from an AVL, is straightforward. It just consists of substituting the smart constructor \texttt{equil} for all the occurrences of the constructor \texttt{Node} in a standard implementation of \texttt{delete} for binary search trees. The reader may consult \cite[Chap. 7]{Pen.05} for more details.

%% file: conclu.tex
\section{Conclusions}
\label{sc:conclu}

The paper has presented the tool Liquid Haskell, and has illustrated its use with a selection of examples of increasing complexity, ranging from preventing pattern matching errors, to proving correct the implementation of AVL-trees. 
Many more examples can be found in the tutorial recently written by the authors and referenced in (\ref{eq:tutorial})(see Sec.~\ref{sc:liquidhaskell}). 
In the preliminary sections, we have presented the Liquid Type technology, of which Liquid Haskell is just an example. 

The combination of {\it recursive refinements} for recursive data types, expressing restrictions on the contents of a data structure (e.g.\ that its elements are sorted), and {\it measures} for defining its structural properties (e.g.\ restrictions on its length, or on its height), gives the system an unexpected big power to express and prove complex properties. The limitations of the Liquid Type approach are those derived of the undecidability of the formula satisfaction problem. If the property being specified needs complex formulas to be proved valid, then the system will give up. For instance, the validity of most universally quantified formulas is undecidable, and these are frequently needed in program verification.

Nevertheless, we believe that this family of systems is worth to be studied because they may change the way in which programmers will think of programs in the future. Rather than following the usual cycle of first write, then compile, then test, and then edit, they could follow a more interesting and profitable one: write type signatures, write code, type-check, and edit. This methodology might drastically lower the number of errors that programmers unadvisedly introduce in programs, without investing much additional effort. The tool is presented as a type-checker, which is already familiar to programmers. So, they might look at it as just a type-checker slightly more evolved  than the usual ones, while what in fact is happening under the hood is that they are doing formal verification. This is possible because most of the tedious and routine proving work is done by the system running in the back.